\documentclass[11pt,a4paper]{article}
\usepackage{amsmath,amsthm,amssymb,amsbsy}
\usepackage{graphicx}
\usepackage{subfig}
\usepackage{multicol}

\usepackage{mathtools}
\usepackage{authblk}
\usepackage{xcolor}
\usepackage{tikz} 

\usepackage{algorithm}
\usepackage{algorithmic}

\title{A Diffusion MRI model for axonal damage quantification based on axial diffusivity reduction in axons: a Monte Carlo simulation study}

\author{Nand Sharma\thanks{Corresponding author email : nandsh11@gmail.com}}

\affil{Department of Radiology, Washington University in St. Louis, Missouri, USA }

\begin{document}

\date{}

\maketitle
\begin{abstract}
Axonal damage is the primary pathological correlate of long-term impairment in multiple sclerosis (MS). Previous work has demonstrated a strong, quantitative relationship between decrease in axial diffusivity and axonal damage. In the present work, we develop an extension of diffusion basis spectrum imaging (DBSI) which can be used to quantify the fraction of diseased and healthy axons based on reduction in axial diffusivity in axons. In this novel method, we model the MRI signal with the axial diffusion (AD) spectrum for each fiber orientation and use two component restricted anisotropic diffusion spectrum (RADS) to model the anisotropic component of the diffusion-weighted MRI signal. Bayesian information criterion (BIC) and model-fitting residuals are calculated to quantify model complexity and goodness of fit. Diffusion coefficients and signal fractions are computed for the optimal model with the lowest BIC score. This gives us the fractions of diseased and healthy axons. We test our method using Monte-Carlo (MC) simulations with the MC simulation package developed as part of this work. The simulation geometry for the voxel includes uniformly spaced cylinders to model axons, and uniformly spaced spheres to model extra-axonal cells. First we test and validate our MC simulations for the basic RADS model. It accurately recovers the fiber and cell fractions simulated, as well as the simulated diffusivities. For testing and validating RADS to quantify axonal damage, we simulate different fractions of diseased and healthy axons. Our method produces highly accurate quantification of diseased and healthy axons with Pearson's correlation (predicted vs true proportion) of $ r = 0.98 $ (p-value = 0.001); the one Sample t-test for proportion error gives the mean error of 2\% (p-value = 0.034). Furthermore, the method recovers the axial diffusivities of the diseased and healthy axons very accurately with mean error of 4\% (p-value = 0.001).

\end{abstract}

\section{Introduction}

Multiple sclerosis (MS) is the most common neurological disorder among young adults with a lifetime risk of one in 400 (\cite{MS_basic}, \cite{MS_basic1}, \cite{MS_basic2}). MS starts in young adults (mean age of onset, 20-30 years), and can progressively lead to physical disability and cognitive impairment (\cite{MS_basic3}). MS affects nearly 2.3 million people worldwide, and nearly 1 million adults in the United States (\cite{MS_prevelance}).
Primarily a neuro-degenerative disease of the central nervous system (CNS), MS is characterized by inflammatory demyelination and axonal damage that results in progressive neurological disabilities (\cite{MS_basic}, \cite{MS_basic1}, \cite{MS_basic2}, \cite{MS_basic3}). 

Axonal damage has been shown to be the main pathological correlate of progressive neurological impairment in MS. The results in \cite{axial1} confirm a strong relationship between reduced axial diffusivity and axonal damage. Diffusion tensor imaging (DTI) has been used previously to detect and distinguish axon and myelin injuries in MS based on axial diffusivity and radial diffusivity respectively (\cite{SONG20031714}, \cite{KIM2006626}, \cite{SUN20061195}, \cite{SUN2}, \cite{KLAWITER20111454}, \cite{doi:10.1212/WNL.0b013e3181e042c4}, \cite{doi:10.1212/WNL.0b013e3181e0434d}). In particular, decreased apparent axial diffusivity and increased apparent radial diffusivity from DTI have been shown to correlate with axonal and myelin damage respectively. But there are limitations of the DTI model due to which it can not distinguish the effects of inflammation-associated vasogenic edema or increased cellularity and the axonal damage. 

DTI is accurate only when there are well-defined tightly packed white matter tracts, as it is confounded by inflammation or crossing fibers (\cite{KLAWITER20111454}). 
To address these limitations of DTI based CNS imaging modalities, more advanced methods such as diffusion basis spectrum imaging (DBSI) (\cite{Wang2011QuantificationOI}) have been used. These multicompartment methods are capable of distinguishing the  diffusion MRI properties associated with axon and myelin injury as well as inflammation in neuro-degenerative diseases (\cite{Lakhani751}, \cite{Sune655}, \cite{axonalLossMSdbsi}). 

The ability of DBSI to isolate the signal contributions from different tissue compartments in CNS and its success in differentiating signal contributions from crossing fibers makes it an ideal method to build on to quantify axonal damage in MS (\cite{Lakhani751}, \cite{Sune655}, \cite{axonalLossMSdbsi}). DBSI models axonal diffusion with anisotropic diffusion tensors, and diffusion in the extra-axonal cells and extracellular space as isotropic diffusion tensors to quantify axonal injury, demyelination and inflammation in the CNS (\cite{Wang2011QuantificationOI},\cite{Lakhani751},\cite{Sune655}, \cite{axonalLossMSdbsi}, \cite{doi:10.1177/1352458518786072}, \cite{CHIANG2014310}).

In the present work, we develop a variation and extension of DBSI which can be used to quantify the fraction of diseased and healthy axons based on the axial diffusivities of the axons in disease and in health. In this method, we first separate the anisotropic component of the diffusion-weighted signal that comes from the axons, and then split it into two components- one coming from healthy axons and the other from the damaged axons. This method- that we call RADS (restricted anisotropic diffusion spectrum) MRI model- is a novel method that uses axial diffusivity reduction as a potential non-invasive biomarker for quantifying the fraction of diseased axons. To the best of our knowledge, no existing method uses axial diffusivity of the axons to distinguish healthy and unhealthy axons.

\section{Background}

In the case of anisotropic diffusion, the dependence of the normalized diffusion-weighted MRI signal $s$ on the diffusion tensor in a pulsed gradient spin echo experiment is given by the relation (\cite{basser1}) :

\begin{equation}
s = \dfrac{s(b)}{s_0} = e^{-b. \bold{r^T} \mathcal{D} \bold{r}},
\end{equation}

where \textbf{r} is a unit column vector in the direction of the diffusion weighting gradient pulse, $\mathcal{D}$ is the diffusion tensor, $s_0$ is the signal without  diffusion weighting, whereas $s(b)$ is the signal with diffusion weighting. The $b$ value is defined as :

$$
b = \gamma^2 G^2 \delta^2 \left(  \Delta-\dfrac{\delta}{3} \right), 
$$
where $\gamma$ is the gyromagnetic ratio of hydrogen proton, $G$ and $\delta$ are the amplitude and duration of the diffusion gradient pulses, and $\Delta$ is the time between the leading edges of the gradient pulses.

Following Anderson's work in \cite{Anderson}, making the assumption that the diffusion tensor is axially symmetric, it can be diagonalized and written as 
\begin{align*}
\bold{D} = \begin{pmatrix} 
            \lambda_{\perp} & 0& 0\\ 
            0 & \lambda_{\perp} &0\\ 
            0 & 0 & \lambda_{\parallel}\\ 
         \end{pmatrix}\,
\end{align*}
in some coordinate system. Here $\lambda_{\perp}$ and $\lambda_{\parallel}$ are the radial and axial diffusivities of the axially symmetric tensor. Let $\alpha$ be the polar angle (the angle from the fiber axis) and $\psi$ be the azimuthal angle (relative to an arbitrary reference direction in the plane perpendicular to the fiber axis). Then diffusion in a particular direction, denoted by $D(\alpha$, $\psi)$, is given by

\begin{align}
D(\alpha, \psi) &= 
\begin{bmatrix} \sin \alpha \cos \psi &  \sin \alpha \sin \psi  & \cos \alpha \\ 
\end{bmatrix}  
\begin{bmatrix} 
            \lambda_{\perp} & 0& 0\\ 
            0 & \lambda_{\perp} &0\\ 
            0 & 0 & \lambda_{\parallel}\\ 
         \end{bmatrix}
\begin{bmatrix} \sin \alpha \cos \psi \\  \sin \alpha \sin \psi  \\ \cos \alpha  
\end{bmatrix}\\
&= \lambda_{\perp} + (\lambda_{\parallel} - \lambda_{\perp}) \cos^2 \alpha.  
\label{dbsi_simplification} 
\end{align}  

In Diffusion basis spectrum imaging (DBSI) (\cite{Wang2011QuantificationOI}), the MRI signal is modeled as a linear combination of the isotropic and anisotropic components. DBSI uses the  simplification in Eq. \ref{dbsi_simplification} of axially symmetric tensor to model the anisotropic component of the signal. To account for multiple orientations of axons, the anisotropic component is modeled as a linear combination of signal contributions from different orientations. As a result, the signal in DBSI is modeled as :

\begin{align} 
s_k = \sum_{i=1}^{N_{ori}} f_{i} e^{-b_k \lambda_{\perp i}} e^{-b_k(\lambda_{\parallel i } - \lambda_{\perp i}) \cos^2\theta_{i k}} 
\notag\\
+ \int_a^b f(D) e^{-b_k D} d(D).
\end{align}
\label{dbsi_mainEq}

In the above Eq. \ref{dbsi_mainEq}, the first term represents the sum of anisotropic components- one for each fiber orientation; $s_k$ is the normalized diffusion-weighted signal corresponding to the $k$th diffusion gradient; $b_k$ is the b-value of the $k$th diffusion gradient; $N_{ori}$ is the number of fiber orientations to be determined; $\theta_{i k}$ is the unknown angle between the $k$th diffusion gradient and the $ith$ fiber direction; $\lambda_{\perp i}$ is the unknown radial diffusivity of the $i$-th anisotropic tensor corresponding to the $i$-th fiber orientation, and $\lambda_{\parallel i }$ is the unknown axial diffusivities in the $i$-th fiber orientation; $f_{i}$ is the unknown signal intensity fraction from the anisotropic tensor corresponding to the $i$-th  fiber orientation; and $D$ is the diffusivity of the isotropic spectrum; and $a$ and $b$ are the low and high diffusivity limits for the isotropic diffusion spectrum $f(D)$.

\section{Axial Diffusion spectrum}
In the present work, we model the MRI signal with the  axial diffusion spectrum for each fiber orientation, not just the mean axial apparent diffusion coefficient (ADC) as is the case with basic DBSI. Furthermore, we assume fibers with the same radial diffusivity for each fiber orientation. Formally,

\begin{align}\label{dbsiAD_mainEq} 
s_k = \sum_{i=1}^{N_{ori}} \sum_{h=1}^{N_{AD}} f_{ih} e^{-b_k \lambda_{\perp i}} e^{-b_k(\lambda_{\parallel i h} - \lambda_{\perp i}) \cos^2\theta_{i k}} 
\notag\\
+ \int_a^b f(D) e^{-b_k D} d(D)
\end{align}

In Eq. \ref{dbsiAD_mainEq}, $s_k$ is the normalized diffusion-weighted signal corresponding to the $k$th diffusion gradient; $b_k$ is the b-value of the $k$th diffusion gradient ($k= 1, 2…N_{grads}$), where $N_{grads}$ is the number of gradient directions; $N_{ori}$ is the number of fiber orientations to be determined; $N_{AD}$ is the number of axial diffusion spectrum components to be determined; $\theta_{i k}$ is the unknown angle between the $k$th diffusion gradient and the $ith$ fiber direction; $\lambda_{\perp i}$ is the unknown radial diffusivity of the $i$-th anisotropic tensor corresponding to the $i$-th fiber orientation, and $\lambda_{\parallel i h}$ are the unknown axial diffusivities in the $h$-th component of the axial diffusion spectrum for the $i$-th fiber orientation; $f_{ih}$ is the unknown signal intensity fraction from the anisotropic tensor corresponding to the $i$-th  fiber orientation and the $h$-th component of the axial diffusion spectrum; and $D$ is the diffusivity of the isotropic spectrum; and $a$ and $b$ are the low and high diffusivity limits for the isotropic diffusion spectrum $f(D)$.

For this work, we consider the case of a single fiber orientation, so that $N_{ori}=1$ and the corresponding radial diffusivity is $\lambda_{\perp}$. With these simplifications, the model in equation \ref{dbsiAD_mainEq} reduces to - 

\begin{align}\label{dbsiAD_reducedModelEq}
s_k=  \sum_{i=1}^{N_{AD}} f_{i} e^{-b_k \lambda_{\perp}} e^{-b_k(\lambda_{\parallel i} - \lambda_{\perp}) \cos^2\theta_{ik}} 
\notag\\
+ \int_a^b f(D) e^{-b_k D} d(D),
\end{align}

where $f_{i}$ is the unknown signal intensity fraction from the anisotropic tensor corresponding to the $i$-th component of the axial diffusion spectrum; $\lambda_{\parallel i}$ are the unknown axial diffusivities in the axial diffusion spectrum; $\theta_{ik}$ is the unknown angle between the $k$th diffusion gradient and the fiber direction.

\subsection{Model Selection}
For model selection as in Eq. \ref{dbsiAD_reducedModelEq}, we solve a non-negative least squares problem. Choosing uniformly spaced $n_1$ points between $0$ and $3 \times 10^{-3} mm^2/s$ for $\lambda_{\parallel}$ and $n_2$ discreet points between $0$ and $3 \times 10^{-3} mm^2/s$ for $D$, and a fixed value for $\lambda_{\perp}$, we form the matrix $\mathcal{M} \in \mathbb{R}^{N\times{(n_1+n_2)}}$ as follows.  

\begin{align} 
\mathcal{M} = \begin{bmatrix} m_{1,1}&. & .& m_{1,n_1} & p_{1,1} &. & .& p_{1,n_2}    \\ 
m_{2,1}&. & . & m_{2,n_1} & p_{2,1} &. & . & .  \\ 
.&. & .& . &. & .  \\ 
&. & .& . &. & .  \\ 
&. & .& . &. & .& .  \\ 
m_{N,1}&. & .& m_{N,n_1} & p_{N,1}  &.& . & p_{N,n_2} \\ 
\end{bmatrix} 
\end{align}

where

\begin{align} 
m_{k, i} = e^{-b_k \lambda_{\perp}} e^{-b_k(\lambda_{\parallel i} - \lambda_{\perp}) \cos^2\theta_{i k}},
\end{align}
and
\begin{align} 
p_{k, i} = e^{-b_k D_i}.
\end{align}

With the above formulation, the model in Eq \ref{dbsiAD_reducedModelEq} can be rewritten in the discretized form as -
\begin{align}\label{discrModel1} 
s_k =  \sum_{i=1}^{n_1} f_{i} m_{ki}
+\sum_{j=1}^{n_2} g_{j} p_{kj},\\
\end{align}
with the constraint 
$$
\sum_{i=1}^{n_1} f_{i} 
+\sum_{j=1}^{n_2} g_{j} = 1.
$$
In matrix form, the signal can be written as follows - 
\begin{equation}
\begin{bmatrix} s_1\\ \dot{•} \\ \dot{•} \\ \dot{•} \\ s_N \end{bmatrix} 
= \begin{bmatrix} m_{1,1}&. & .& m_{1,n_1} & p_{1,1} &. & .& p_{1,n_2}    \\ 
m_{2,1}&. & . & m_{2,n_1} & p_{2,1} &. & . & .  \\ 
.&. & .& . &. & .  \\ 
&. & .& . &. & .  \\ 
&. & .& . &. & .& .  \\ 
m_{N,1}&. & .& m_{N,n_1} & p_{N,1}  &.& . & p_{N,n_2} \\ 
\end{bmatrix}  
\begin{bmatrix} f_1\\ \dot{•} \\ \dot{•} \\ \dot{•} \\ \dot{•} \\ \dot{•} \\ f_{(n_1+n_2)} \end{bmatrix}, 
\end{equation}
or, equivalently as
\begin{equation}
\bold{ s = Mf}
\end{equation}
with the constraints
$$
\sum f_{i} = 1, \hspace{.04in} 0 \leq f_i \leq 1 \hspace{.04in} \forall i.
$$
The problem of model selection reduces to an optimization problem where our goal is to find the optimal values of $f_i$, $i=1,2,..(n_1+n_2)$, with the constraints that $f_i$ take non-negative values. This is solved using non-negative least squares with $l2$ regularization as follows. 

\begin{equation}\label{l2min}
\min_{\bold{f}} \| \bold{s-Mf} \| + \beta \|  \bold{f} \|_2, \hspace{.1in} \sum f_{i} = 1, \hspace{.04in} 0 \leq f_i \leq 1 \hspace{.04in} \forall i.
\end{equation}

For uniformly spaced values of $\lambda_{\perp}$ between $0$ and $0.4 \times 10^{-3} mm^2/s$, we solve the minimization problem in Eq. \ref{l2min}, and pick the model with the least sum of squared residuals. This gives us the best fit model with the best fit radial diffusivity for the anisotropic component of the signal and the spectrum for the isotropic components. 

Please observe that although we consider a single fiber orientation in the present case, the same model selection procedure works for multiple fiber orientations. The matrix $\mathcal{M}$ in that case is such that $\mathcal{M} \in \mathbb{R}^{N\times{( (n_1 * N_{ori})+n_2)}}$. In that case the number of parameters increase for the minimization problem in Eq. \ref{l2min}.

%
%
%
%
\subsection{Two component Restricted Anisotropic Diffusion Spectrum (RADS) model for Axonal Health}
After solving the above minimization problem, we split the signal into anisotropic and isotropic components, and consider the anisotropic part of the signal $\bold{s^{an}}$ such that 

\begin{align} 
s_k^{an} =  s_k - \sum_{j=1}^{n_2} g_{j} p_{kj}.
\end{align}

While the isotropic components model the extra-axonal-extra-cellular and intra-cellular spins, the anisotropic component models the intra-axonal spins. For identifying the healthy and diseased axons, the anisotropic part of signal $s^{an}_k$ is modeled as a linear combination of average signal from each of the two components - healthy and diseased axons :

\begin{align}\label{dbsiADS_axonHealthEq}
s_k^{an}=\underbrace{f e^{-b_k \lambda_{\perp}} e^{-b_k(\lambda_{\parallel 1} - \lambda_{\perp}) \cos^2\theta_{1 k}}}_\text{unhealthy axons} +  \underbrace{(1-f) e^{-b_k \lambda_{\perp}} e^{-b_k(\lambda_{\parallel 2} - \lambda_{\perp}) \cos^2\theta_{2 k}}}_\text{healthy axons}
\end{align}

where $f$ is the unknown signal intensity fraction from the anisotropic tensor corresponding to the unhealthy axons, and $(1-f)$ the signal intensity fraction from the anisotropic tensor corresponding to the healthy axons. With $ \bold{M^{an}}$ defined as 

\begin{equation}
\bold{M^{an}}
= \begin{bmatrix} m_{1,1}&. & .& m_{1,n_1}    \\ 
m_{2,1}&. & . & m_{2,n_1}   \\ 
.&. & .& .   \\ 
&. & .& .   \\ 
&. & .& . \\ 
m_{N,1}&. & .& m_{N,n_1}  \\ 
\end{bmatrix}  
= \begin{bmatrix} \bold{M^{an}_1} & \bold{M^{an}_2} & \ldots  \bold{M^{an}_{n_1}} \end{bmatrix}, 
\end{equation}
where the column vector $\bold{M^{an}_{n_1}}$ corresponds to the $\lambda_{\parallel} = 2 \times 10^{-3} mm^2/s$, assumed to be the axial diffusivity for the healthy axons. We solve the minimization problem again by using non-negative least squares for each value of $i$, $(1 \leq i \leq n_1-1)$ :

\begin{equation}\label{RASmin}
\min_{f} \| \bold{s^{an}-M^{an}_i} f - \bold{M^{an}_{n_1}} (1-f) \| , \hspace{.1in} 0 \leq f \leq 1.
\end{equation}

This fitting process varied the $\lambda_{\parallel}$ values for the unhealthy compartment, keeping the $\lambda_{\parallel}$ for the healthy compartment fixed at $2 \times 10^{-3} mm^2/s$; also the $\lambda_{\perp}$ was kept fixed at $0$. This minimization of the sum of squared residuals gives us $(n_1 -1)$ models. We compare these models using Bayesian information criterion (BIC) \cite{bicBase}, and the model with the lowest BIC score is chosen, which is equivalent in this case to the solution with the minimum prediction error. The optimal model gives us the fractions of diseased and healthy axons and the axial diffusivities of the diseased and healthy axons. Again, observe that in the case of crossing fibers, the Eq. \ref{dbsiADS_axonHealthEq} is solved for each fiber orientation, and in that case the use of BIC score becomes important to pick the optimal solution. We decided to use  BIC score for optimal solution in the more general case.

\section{Monte-Carlo Simulations}

Brownian motion of water molecules was examined using Monte-Carlo (MC) simulation on a computer-generated mouse optic nerve model. The tissue model consists of a coherent axon bundle mimicked by cylindrical tubes of $1 \mu m$ radius (anisotropic compartment) of infinite length within imaging echo time ($TE$). This is done such that the fibers make up $35 \%$ of the voxel volume \cite{LAMPINEN2023120338}. The cylindrical tubes are surrounded by randomly distributed extra-axonal cells, modeled by spheres with a $5.3 \mu m$ radius of set volume fraction (5\%) \cite{LAMPINEN2023120338}. 

We tracked $10^6$ water molecules uniformly distributed into the tissue model monitoring the trajectory at $5 \mu s$ time-step of individual molecule undergoing random walk three-dimensionally. We assume that both axon and cell membranes are impermeable, therefore, the water molecules elastically bounce off these membranes. We have used the diffusivity $3 \times 10^{-3} mm^2/s$ for free water and $2 \times 10^{-3} mm^2/s$ for healthy axons \cite{Bonaiuto}. Axonal injury was simulated by setting intra-axonal diffusivity at a reduced value ($1\times 10^{-3} mm^2/s$) for damaged axons. We simulated three levels of injury severity for the current MC simulations.

The diffusion weighted MRI signals are generated by using a simple spin-echo diffusion-weighting sequence. The phase accumulation of each water molecule during echo time ($TE$) was calculated according to the random walk trajectory. The diffusion-weighting parameters for Monte-Carlo simulation were set according to in vivo mouse optic nerve DBSI \cite{Brunsing}: time between gradient pulses ($\Delta$) = 18 ms, gradient pulse duration ($\delta$) = 6 ms, and maximum b-value =$3000 s/mm^2$, and TE in this simulation was set to be $100 ms$. Diffusion encoding was done with $25$ icosahedral directions with one extra non-diffusion weighed image. The $T1$ and $T2$ relaxation effects were not considered. Each simulation condition was repeated 10 times.

\begin{figure}%
    \centering
    \subfloat[Simulation Geometry]{{\includegraphics[width=5cm]{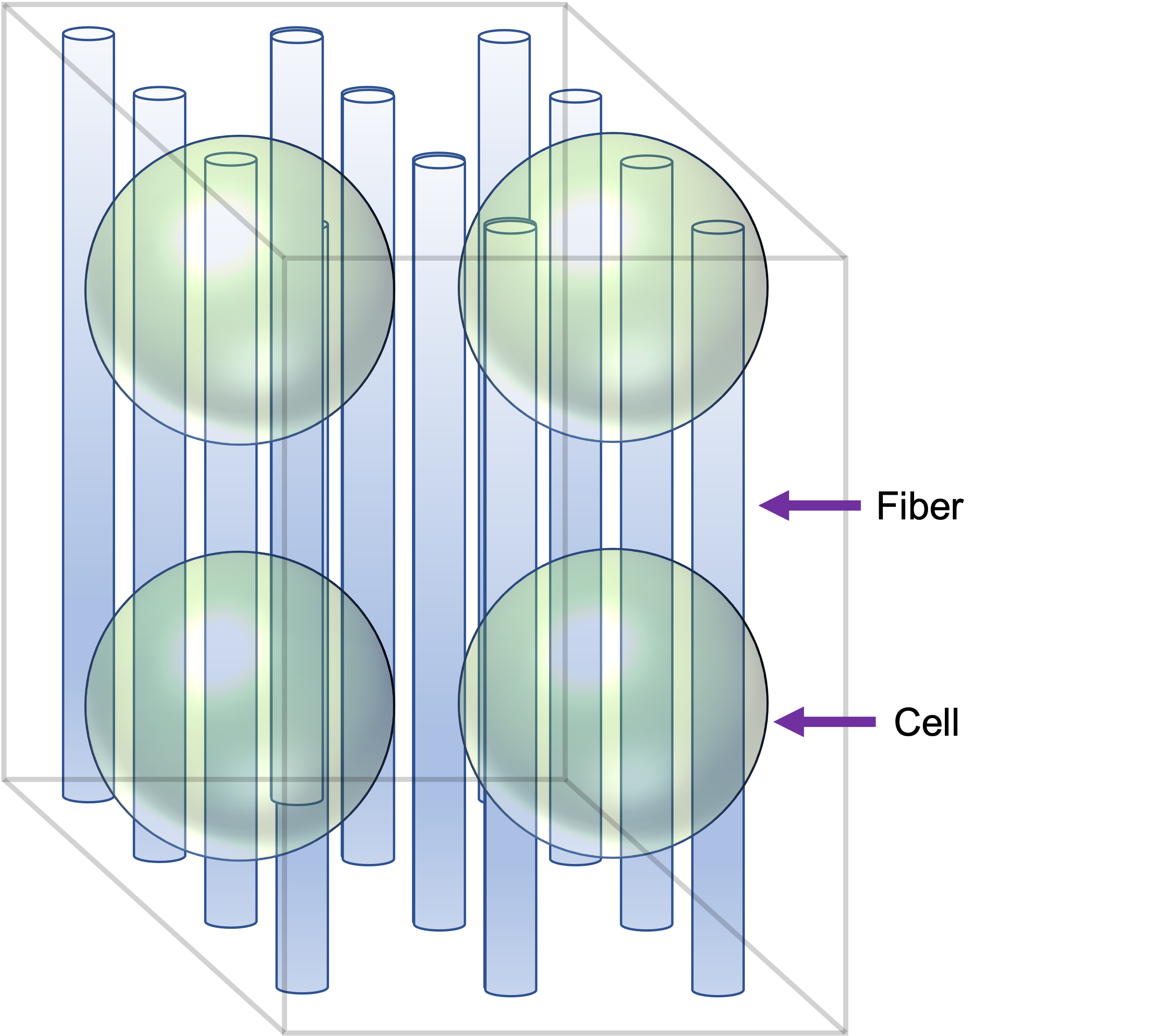} }}%
    \qquad
    \subfloat[ Cross-sectional view]{{\includegraphics[width=6cm]{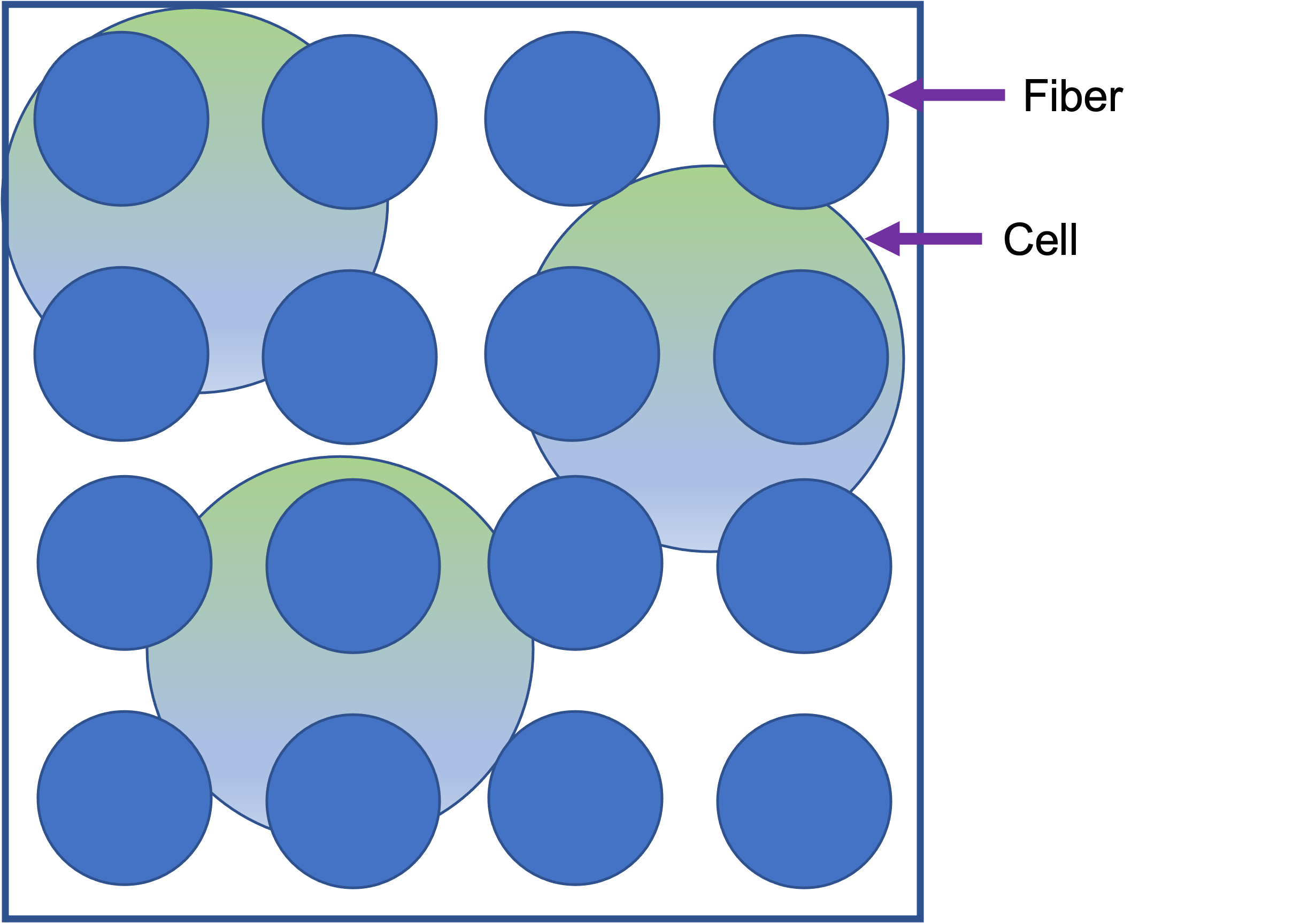} }}%
    \caption{Monte-Carlo simulation geometry. On the left is a voxel with cylindrical axons, and spherical extra-axonal cells.}%
    \label{fig:example}%
\end{figure}

\begin{table}[!ht]
\caption{Simulation parameters} 
\centering 
\begin{tabular}{c c c c} 
\hline\hline 
Parameter & Simulated value \\ [0.5ex] 
\hline 
$\delta$ & 6 ms   \\
$\Delta$ & 18 ms   \\
Timestep & 5 $\mu $s   \\
TE & 100 ms\\
Voxel side & 100$\mu $m  \\
Fibers volume & $35\%$  \\ 
Cell volume & $5\%$  \\
Fiber radius & $1-3 \mu $m  \\ 
Cell radius & $5-10 \mu $m  \\ 
No. of spins & $1,000,000$  \\
b-values & $0-3000$ $ s/mm^2$\\
No. of Gradient directions & 26\\
[1ex] 
\hline 
\end{tabular}
\label{table:simParams} 
\end{table} 
With our voxel geometry (Figure \ref{fig:example}), we get a cell fraction of $5\%$ with the spheres of $5.3 \mu m$ radius. We have three compartments in a voxel with different diffusion dynamics - Intra-axonal (IA), Intra-cellular-extra-axonal (ICEA), and extra-axonal-extra-cellular (EAEC). In the ICEA and EAEC compartments, the diffusivity is taken to be the free water diffusivity i.e. $3 \times 10^{-3} mm^2/s$, whereas the diffusivity in the IA compartment is assumed to be $2 \times 10^{-3} mm^2/s$ for healthy axons.

\begin{figure}[!ht]%
    \centering
    \subfloat[\centering Spin trajectories in the three compartments]{{\includegraphics[width=4in]{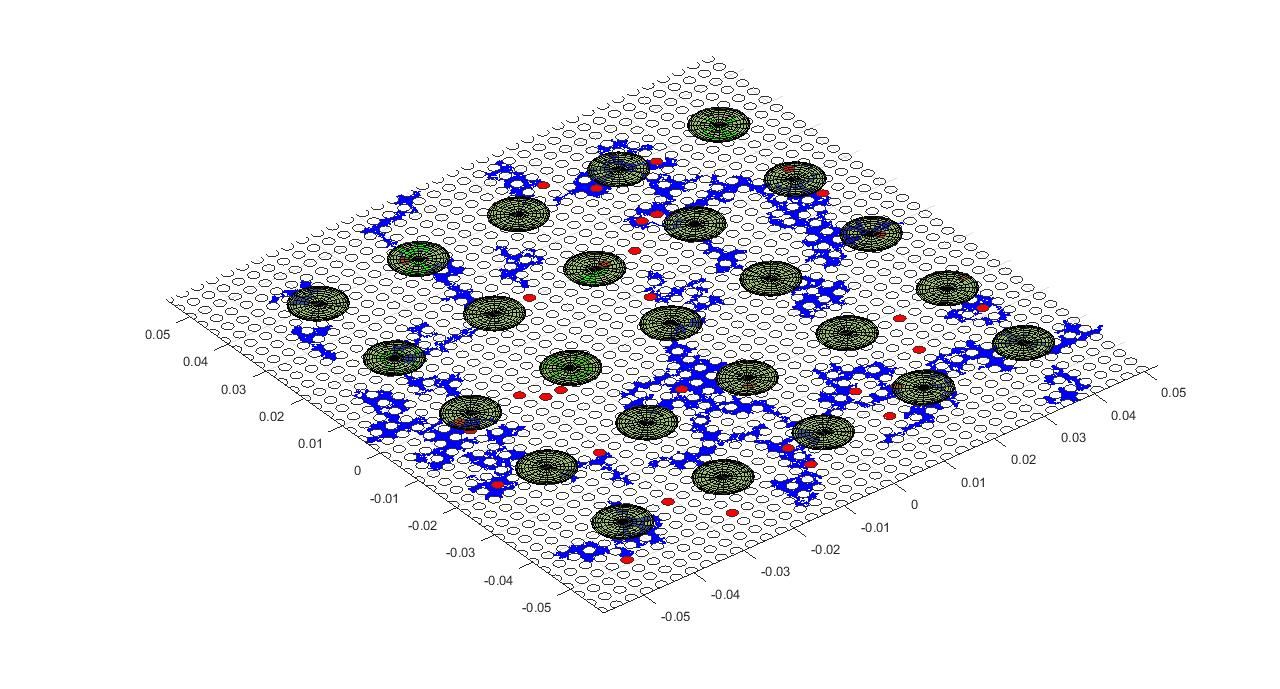} }}%
    \qquad
    \subfloat[\centering Spin trajectory in the Fiber compartment]{{\includegraphics[width=2in]{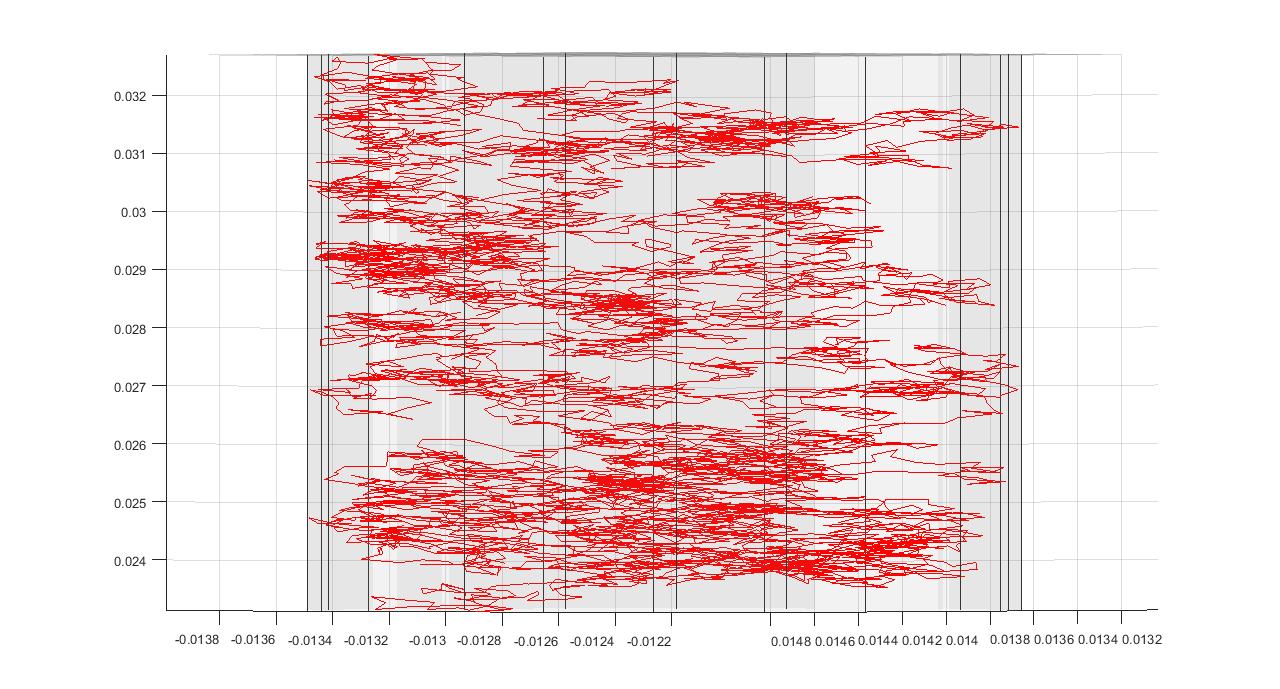} }}%
     \subfloat[\centering Spin trajectory in the Cell compartment]{{\includegraphics[width=2in]{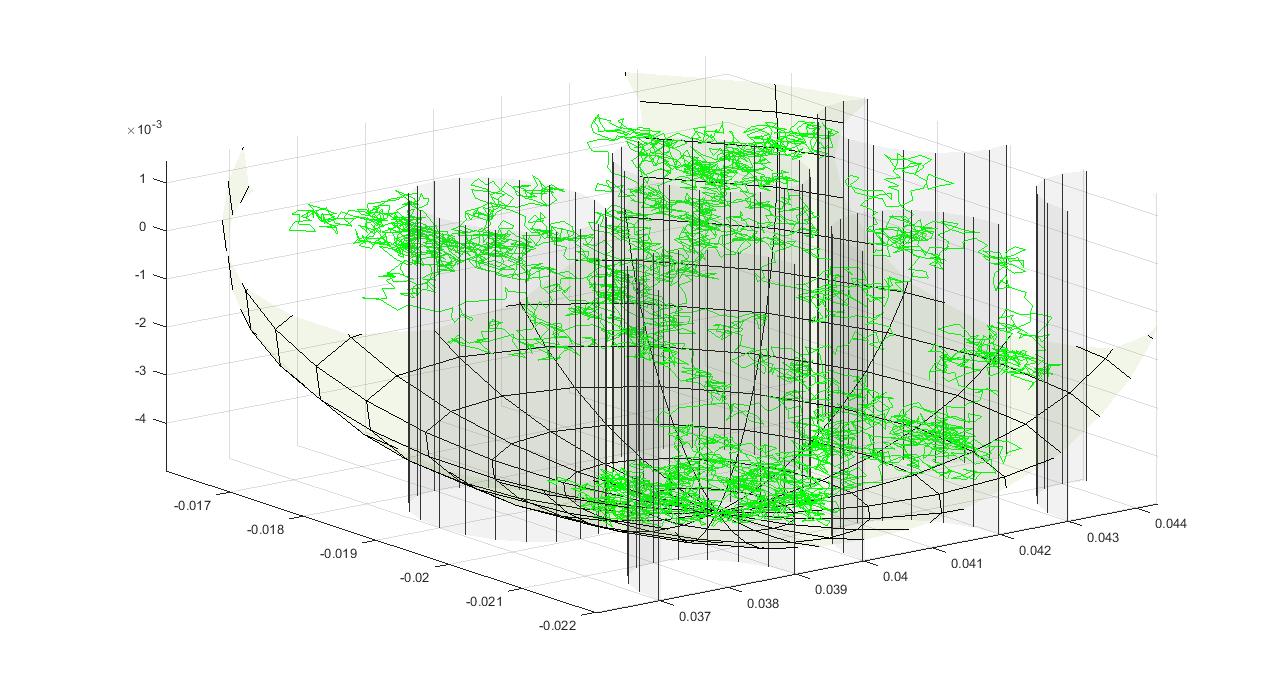} }}%
    \qquad
    \subfloat[\centering Spin trajectory in the free compartment]{{\includegraphics[width=2in]{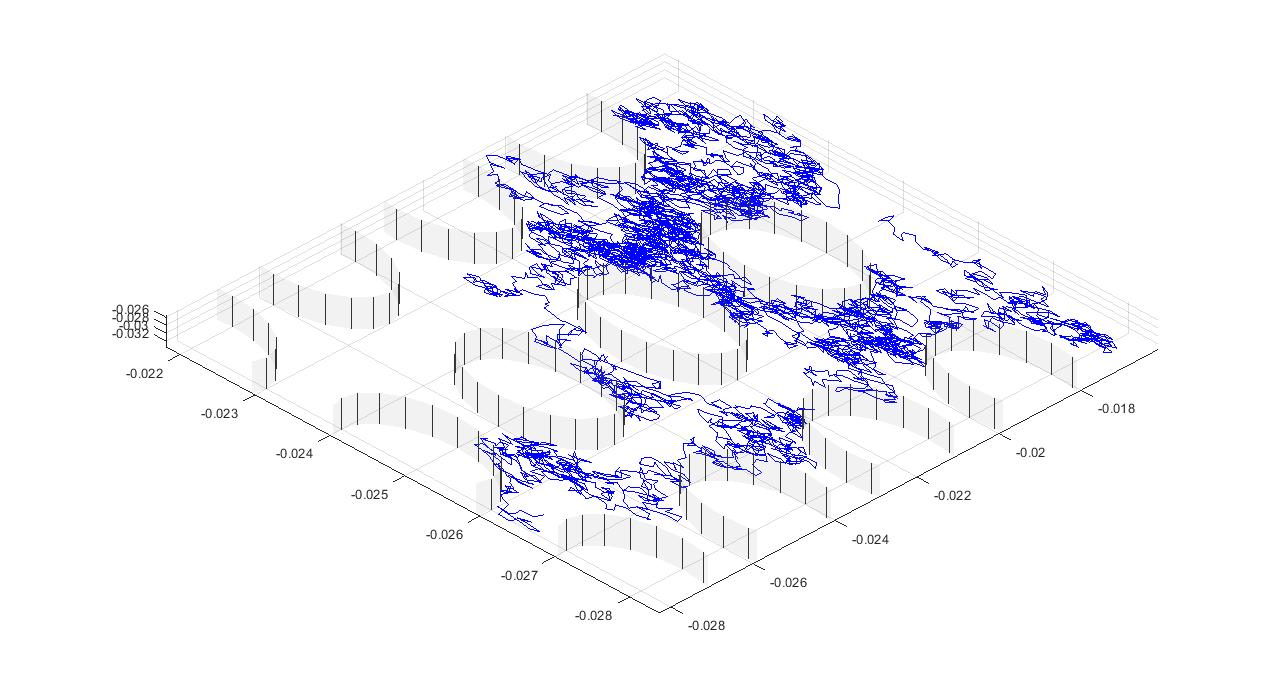} }}%
    \caption{Example Spin trajectories in MC simulations}%
    \label{fig:trajectories}%
\end{figure}

The number of spins ($10^6$) is arrived at after detailed analysis of the number of spins required to stably and accurately produce the expected mean diffusivities as simulated. Recall that, in one dimensional case, the squared displacement of the particles from their starting point over a time $\tau$ in free unrestricted diffusion (isotropic diffusion), averaged over all the sampled particles, is directly proportional to the observation time (\cite{Einstein}). This is denoted as the Einstein Equation : 
\begin{align}
\left\langle r^2 \right\rangle = 2D\tau
\label{EinsteinEq}
\end{align}

where $\left\langle.\right\rangle$ accounts for the averaging operation, $r$ for the displacement and the scalar constant $D$, known as the self diffusion coefficient, measures the mobility of the particle ensemble. In the case of anisotropic diffusion, the analogous model is the diffusion tensor (DT) model proposed by Basser et al. (\cite{Basser}). Here the scalar diffusion coefficient D is replaced by a positive symmetric semi-definite matrix $\mathcal{D}$ representing diffusion, the diffusion tensor. Therefore, Einstein’s relation (Eq. \ref{EinsteinEq}) is generalized, relating the covariance matrix of the net displacement vector $\textbf{R} \in \mathbb{R}^3$ and diffusion tensor $\mathcal{D}$ as

\begin{align*}
\mathcal{D} = \begin{pmatrix} 
            \mathcal{D}_{xx} & \mathcal{D}_{xy} & \mathcal{D}_{xz}\\ 
            \mathcal{D}_{xy} & \mathcal{D}_{yy} & \mathcal{D}_{yz}\\ 
            \mathcal{D}_{xz} & \mathcal{D}_{yz} & \mathcal{D}_{zz}\\ 
         \end{pmatrix}\,
         = \dfrac{1}{6\tau} \left\langle \textbf{R} \textbf{R}^T \right\rangle.
\end{align*}

Diagonalization of diffusion tensor $\mathcal{D}$ gives the axial and radial diffusivities for the cylindrical IA compartment; for ICEA and EAEC compartments, the diagnolization gives the diffusivities in the $x,y,z$ directions. We use this diagnolization after using net displacements to find the number of spins that gives stable results in multiple simulation runs. We found that $10^6$ spins produce the required accuracy and acceptable variance.


Each spin takes a step in a random direction in the 3D space. This is achieved by using random 3D vector drawn from a normal 3D distribution, which is an effective and low cost method to get points uniformly spaced on a unit sphere, which gives uniformly distributed directions. The step size is taken to be $\sqrt{ 6D t_s}$ where  $t_s$ is the fixed time step, and $D$ is the diffusion coefficient. We use a time step of $5 \mu s$ for our experiments. The detailed simulation parameters for the experiments are shown in Table \ref{table:simParams}.  The Figure \ref{fig:trajectories} shows a few spin trajectories in the three different compartments.



\subsection{PGSE Stejskal-Tanner sequence} 
\begin{figure}[!h]
\begin{center}
\includegraphics[width=4in]{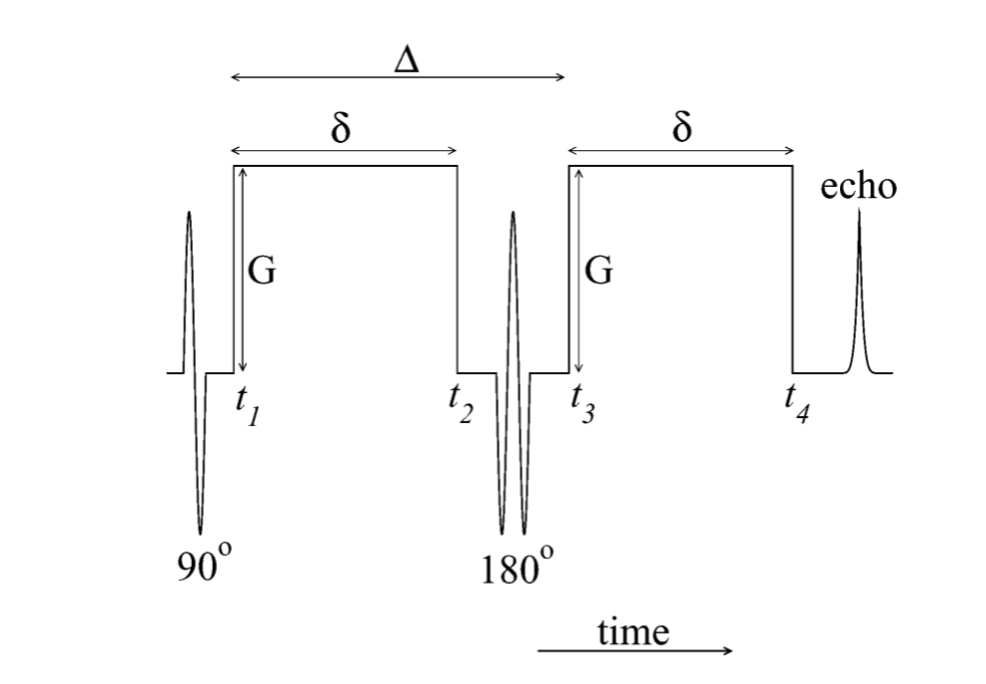}
\end{center}
\caption{Pulsed gradient spin-echo (PGSE) Stejskal-Tanner sequence.}
\label{PulseSeq}
\end{figure}

The schematic of a Pulsed gradient spin-echo (PGSE) Stejskal-Tanner sequence \cite{stejskal1965spin} \cite{Vlaardingerbroek2003}, used for simulations is shown in Figure \ref{PulseSeq}.

During the first half of the pulse sequence, a spin will accumulate a phase shift given by
\begin{equation}
\phi_1(\textbf{G},\delta) = \gamma \textbf{B}_{0} +\gamma \int_{t_1}^{t_2}
\textbf{G}.\textbf{x}(t) dt,
\end{equation}

where $\gamma$ is the proton gyromagnetic ratio (42 MHz/Tesla), $\textbf{G}$ is the strength of the diffusion sensitizing gradient pulses, $\delta$ is the duration of the diffusion gradient pulses, and $\Delta = t_3 - t_1$ is the time between diffusion gradient RF pulses. The phase shift experienced by each spin in the second half of the pulse sequence is given by
\begin{equation}
\phi_2(\textbf{G},\delta) = \gamma \textbf{B}_{0} +\gamma \int_{t_3}^{t_4}
\textbf{G}.\textbf{x}(t) dt
\end{equation}
The net phase shift experienced by a spin is the difference between the two phase shifts - 

\begin{equation}
\phi(\textbf{G},\delta) = \phi_1(\textbf{G},\delta) - \phi_2(\textbf{G},\delta).
\end{equation}

So, the net phase shift experienced by a spin is
\begin{equation}
\phi(\textbf{G},\delta) = \gamma \textbf{G}.\left(  \int_{t_1}^{t_2}
\textbf{x}(t) dt  - \int_{t_3}^{t_4}
\textbf{x}(t) dt \right) 
\end{equation}

For the system with the whole ensemble of spins, the signal is given by

\begin{equation}
S(t=TE) = S_0 \int P(\phi,t) \mathrm{e}^{-i \phi} d \phi
\end{equation}

The produced attenuated signal is then the result of the accumulated phase shift of the full assembly of spins at time TE, given by

\begin{equation}\label{attSignal}
S(t=TE)/{S_0}  = \left\langle   \mathrm{e}^{-i \int_{0}^{TE} \phi(t)dt } \right\rangle. 
\end{equation}

MC simulation uses finite number of spins $N_s$ so that the attenuated signal in Eq. \ref{attSignal} is approximated as 

\begin{equation}\label{attSignalsim}
S(t=TE)/{S_0}  = \dfrac{1}{N_s} \sum_{j=1}^{N_s} \mathrm{e}^{-i \sum_{1}^{N_t} \phi_{jk}(t)t_s }, 
\end{equation}

where $N_t$ is the number of steps in the random walk, $\phi_{jk}$ is the phase shift of spin $j$ in step $k$, and $t_s$ is the time step.

We use fixed time step $t_s$, related to step length $l_s$ as
\begin{equation}
l_s = \sqrt{ 6D t_s}
\end{equation}
where $D$ is the diffusion coefficient. After the spin trajectories are simulated, the signal is computed using Eq. \ref{attSignalsim}.

\section{MC Simulation Experiments}

Before simulating the axonal injury, we simulate and validate our model with simpler cases. These cases simulate a single average diffusivity for the random walk in all three compartments - the IA, the EAIC and the EAEC  compartments. This is a useful first step to see what ADCs are observed for the isotropic components, and what axial and radial diffusivities are observed in the anisotropic part. We use the simple geometry of our voxel structure, to better control and observe the dynamics of the spins and how well our MC simulations and the model perform. To that end, as the first set of models, we consider only three directions diffusion weighting- $x,y,z$. Observe that in the $z$-direction, $\theta_{i k} = 0$, so the overall model in Eq. \ref{dbsiAD_reducedModelEq} reduces to :

\begin{figure}[!htb]
\subfloat[\centering ]{{\includegraphics[width=2in]{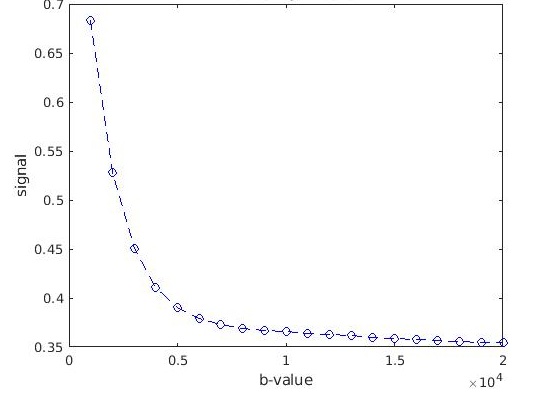}}}%
\subfloat[\centering ]{{\includegraphics[width=2in]{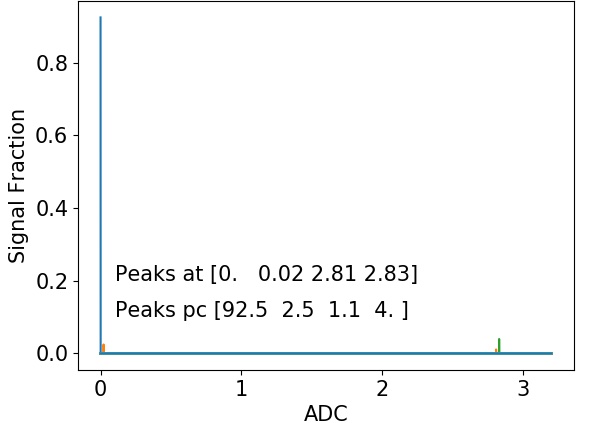}}}%
\qquad
\subfloat[\centering]{{\includegraphics[width=2in]{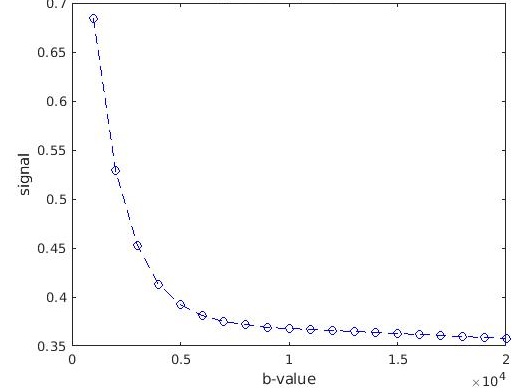}}}
\subfloat[\centering]{{\includegraphics[width=2in]{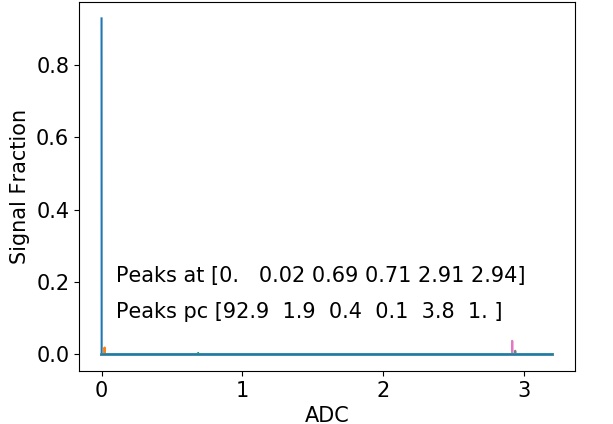}}}%
\qquad
\subfloat[\centering]{{\includegraphics[width=1.9in]{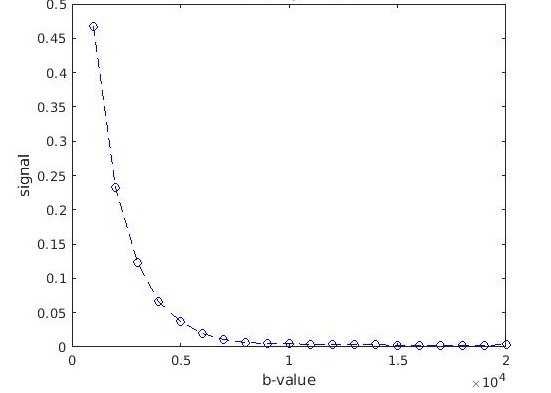}}}
\subfloat[\centering ]
{{\includegraphics[width=2in]{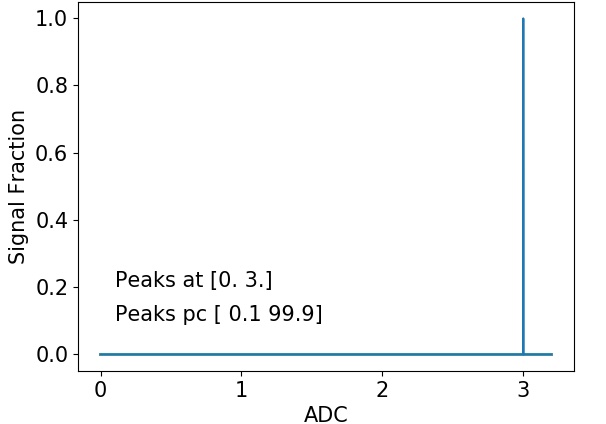}}}%
\caption{Fiber Compartment Signal and Signal Intensities (the units for the ADC are $mm^2/s$ and $s/mm^2$ for the b-values). The top row shows the signal decay with increasing b-values, and the predicted ADC and signal fraction with diffusion weighting in the $x$-direction. The middle row shows the same for diffusion weighting in the $y$-direction; and the bottom shows the graphs for the diffusion weighting in the $z$-direction.}
\label{fiberComp1}%
\end{figure}

\begin{figure}[!htb]
\subfloat[\centering ]{{\includegraphics[width=2in]{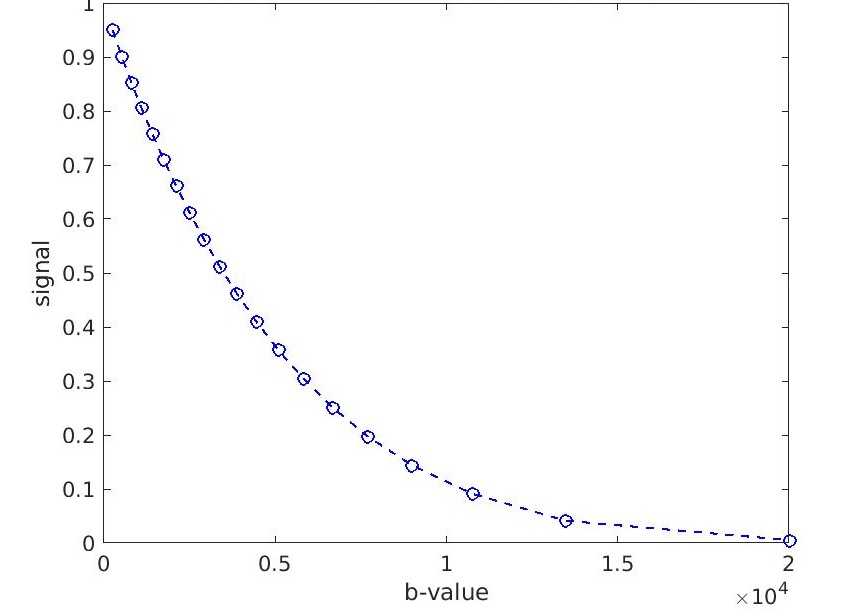}}}%
\subfloat[\centering ]{{\includegraphics[width=2in]{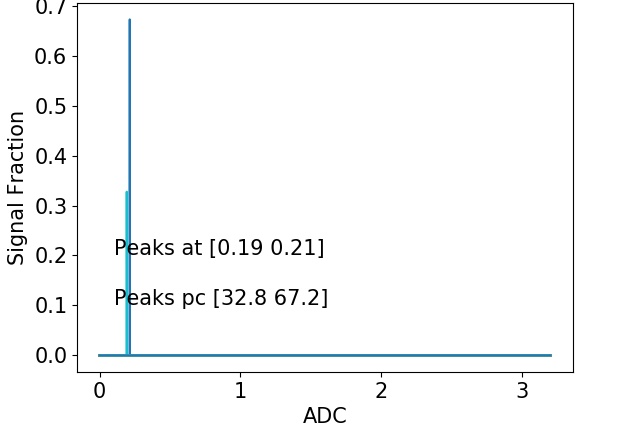}}}%
\qquad
\subfloat[\centering ]{{\includegraphics[width=2in]{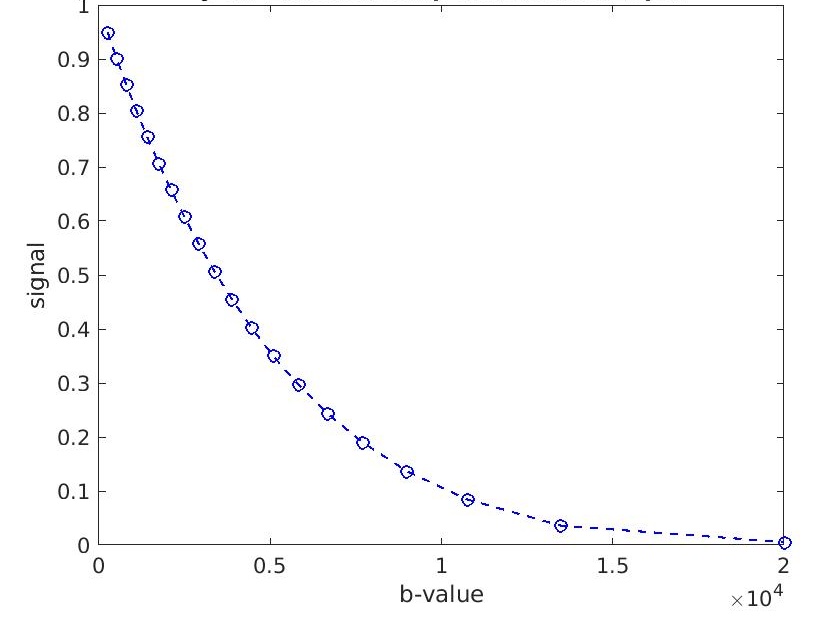}}}%
\subfloat[\centering ]{{\includegraphics[width=2in]{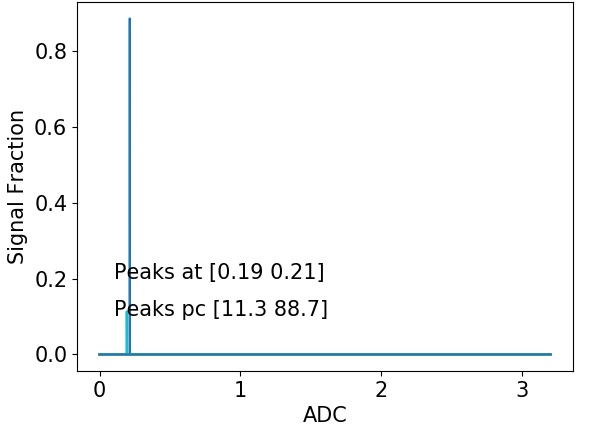}}}
\qquad
\subfloat[\centering ]{{\includegraphics[width=2in]{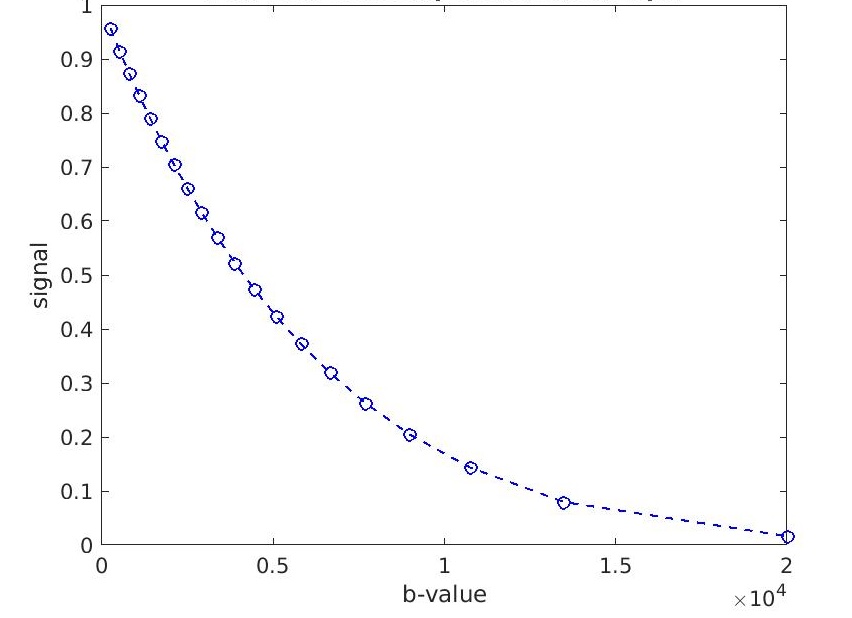}}}%
\subfloat[\centering ]{{\includegraphics[width=2in]{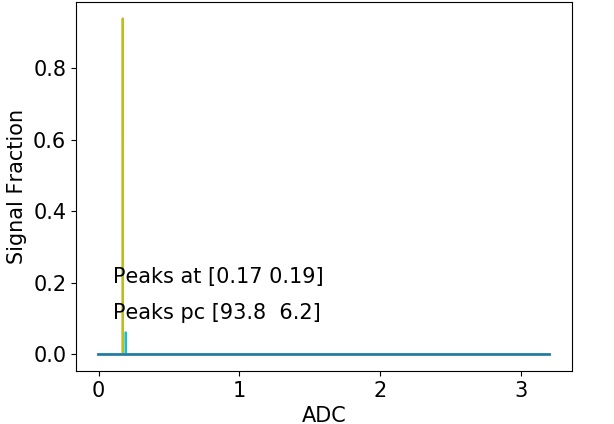}}}%
\caption{Cell Compartment Signal and Signal Intensities(the units for the ADC are $mm^2/s$ and $s/mm^2$ for the b-values). The top row shows the signal decay with increasing b-values, and the predicted ADC and signal fraction with diffusion weighting in the $x$-direction. The middle row shows the same for diffusion weighting in the $y$-direction; and the bottom shows the graphs for the diffusion weighting in the $z$-direction.}
\label{cellComp1}%
\end{figure}

\begin{figure}[!htb]
\subfloat[\centering ]{{\includegraphics[width=2in]{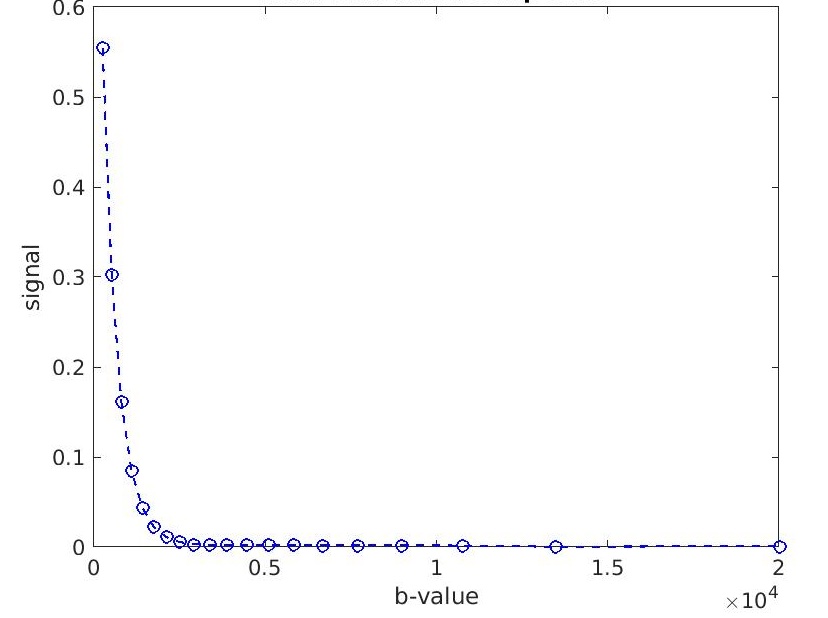}}}%
\subfloat[\centering ]{{\includegraphics[width=2.2in]{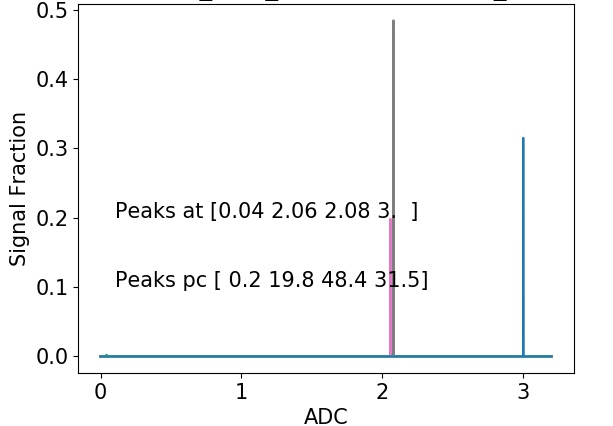}}}%
\qquad
\subfloat[\centering ]{{\includegraphics[width=2in]{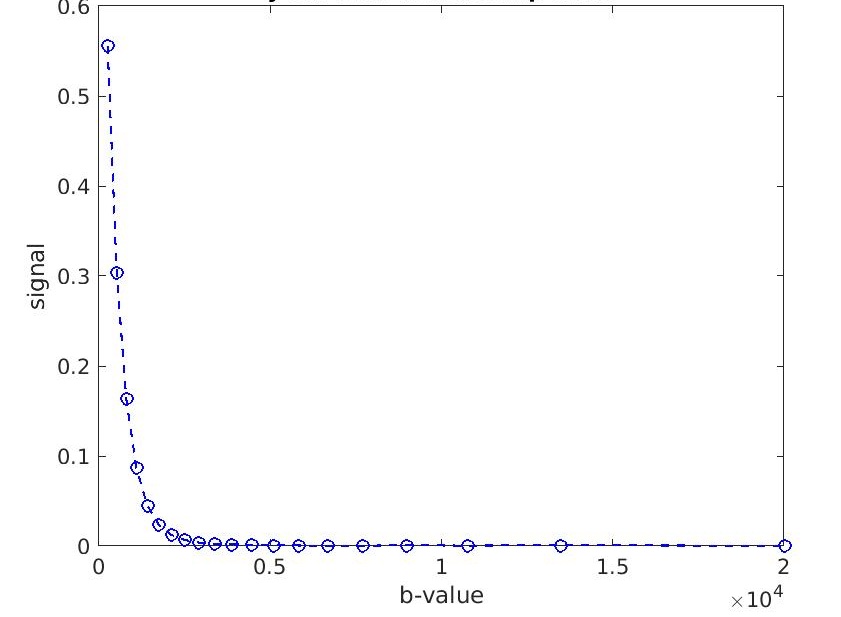}}}%
\subfloat[\centering ]{{\includegraphics[width=2.2in]{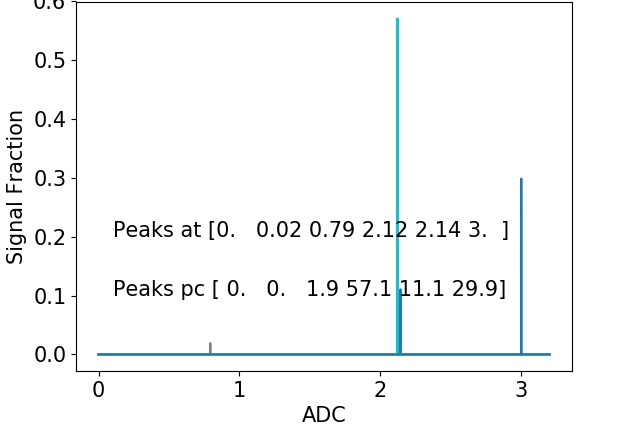}}}%
\qquad
\subfloat[\centering ]{{\includegraphics[width=2in]{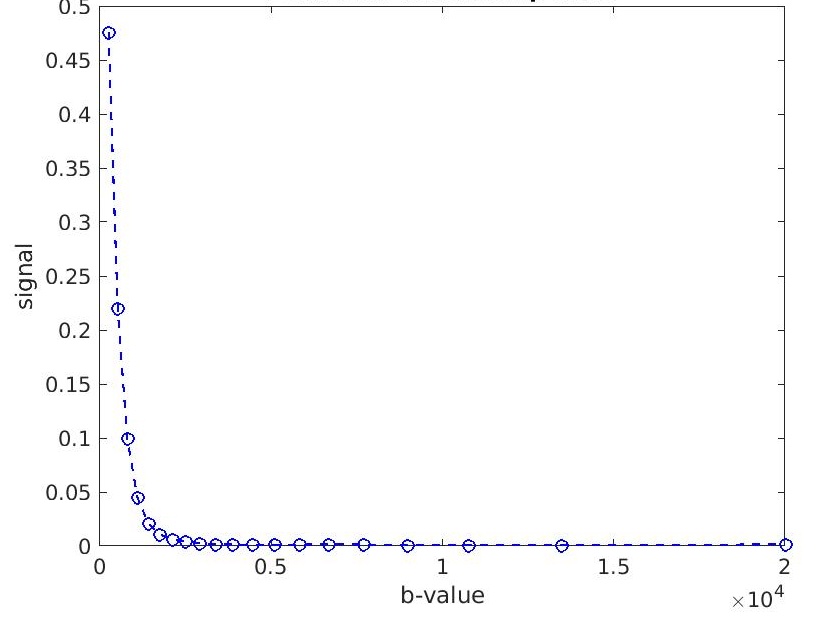}}}%
\subfloat[\centering ]{{\includegraphics[width=2.2in]{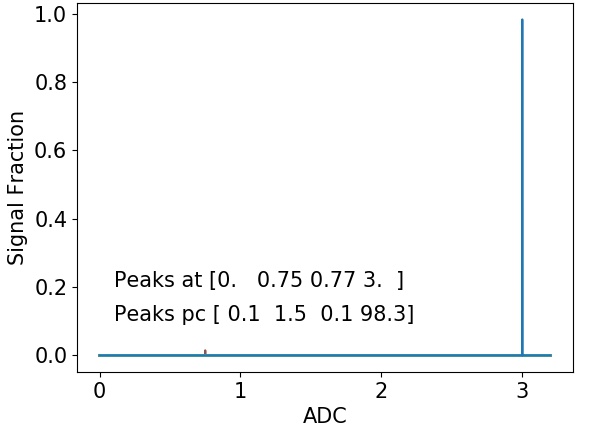}}}%
\caption{Free Water Compartment Signal and Signal Intensities (the units for the ADC are $mm^2/s$ and $s/mm^2$ for the b-values). The top row shows the signal decay with increasing b-values, and the predicted ADC and signal fraction with diffusion weighting in the $x$-direction. The middle row shows the same for diffusion weighting in the $y$-direction; and the bottom shows the graphs for the diffusion weighting in the $z$-direction.}
\label{waterComp1}%
\end{figure}


\begin{figure}[!htb]
\subfloat[\centering ]{{\includegraphics[width=2in]{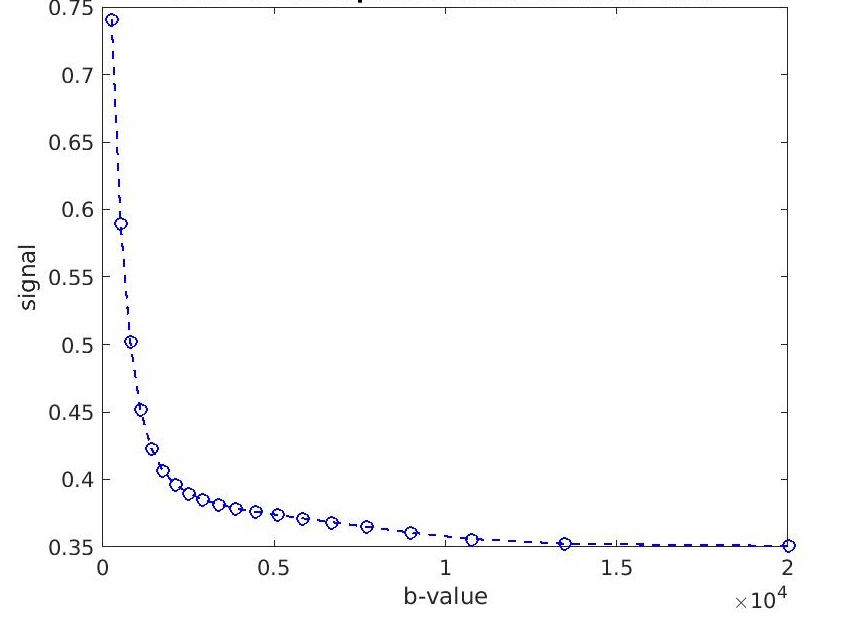}}}%
\subfloat[\centering ]{{\includegraphics[width=2.2in]{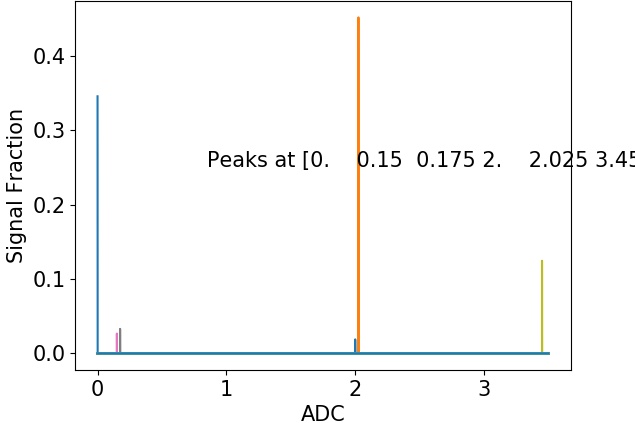}}}%
\qquad
\subfloat[\centering ]{{\includegraphics[width=2in]{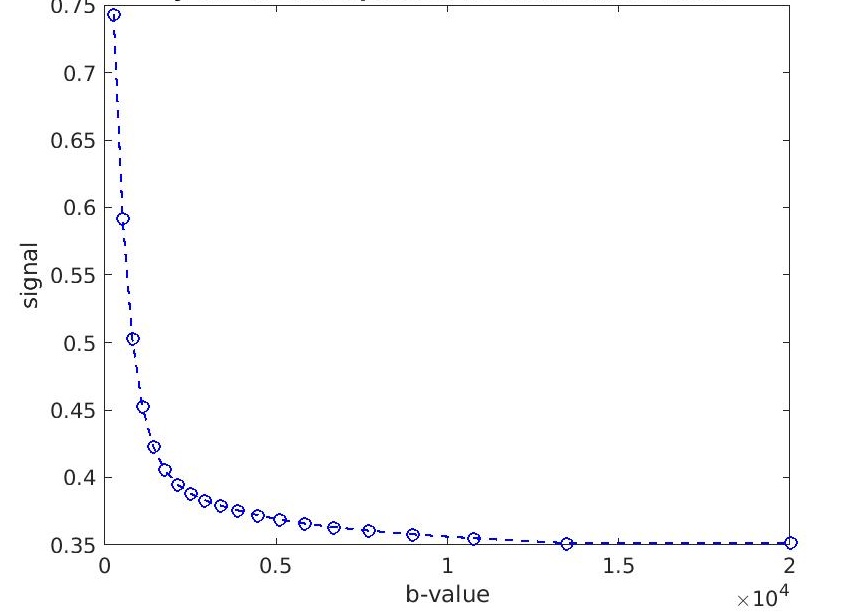}}}%
\subfloat[\centering ]{{\includegraphics[width=2.2in]{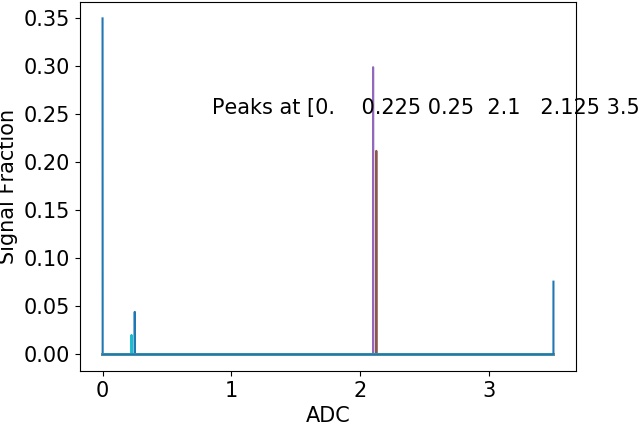}}}%
\qquad
\subfloat[\centering ]{{\includegraphics[width=2in]{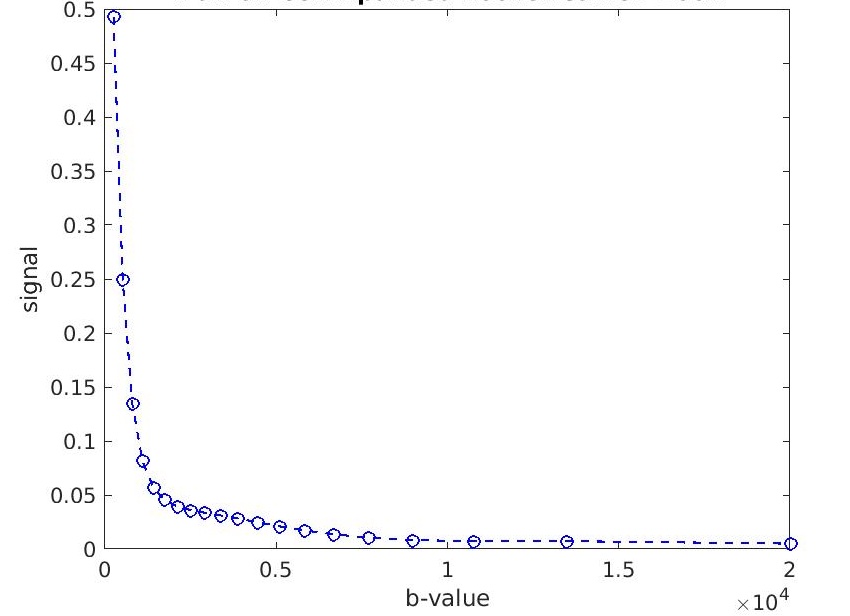}}}%
\subfloat[\centering ]{{\includegraphics[width=2.2in]{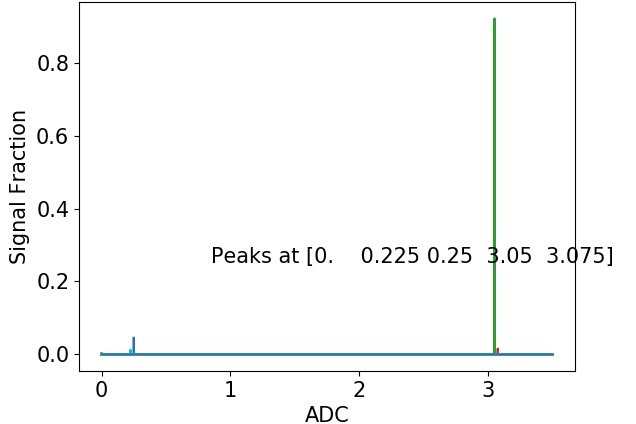}}}%
\caption{Full Structure Signal and Signal Intensities (the units for the ADC are $mm^2/s$ and $s/mm^2$ for the b-values). The top row shows the signal decay with increasing b-values, and the predicted ADC and signal fraction with diffusion weighting in the $x$-direction. The middle row shows the same for diffusion weighting in the $y$-direction; and the bottom shows the graphs for the diffusion weighting in the $z$-direction. }\label{fullStr1}
\end{figure}

\begin{align}\label{dbsiAD_reducedModel_z} 
s_k &=  f_1 e^{-b_k \lambda_{\perp}} e^{-b_k(\lambda_{\parallel i} - \lambda_{\perp}) }
+\sum_{j=2}^{n_2} f_{j} e^{-b_k D}\\
&=  f_1  e^{-b_k(\lambda_{\parallel i}) }
+\sum_{j=2}^{n_2} f_{j} e^{-b_k D}
\end{align}

In the $x$ (and $y$)-direction, $\theta_{i k} = \pi/2$, so the  the overall model in Eq. \ref{dbsiAD_reducedModelEq} reduces to :

\begin{align}\label{dbsiAD_reducedModel_xy} 
s_k &=  f_1 e^{-b_k \lambda_{\perp}} 
+\sum_{j=2}^{n_2} f_{j} e^{-b_k D}
\end{align}
For these simpler cases, we use the $b-$values between $0$ and $2 \times 10^4$ $s/mm^2$, and the three coordinate directions as diffusion weighting directions. Although this range of b-values are not realistic for MRI-scanners, this is done to see that the model and the MC simulations are validated.
For the full simulations for axonal damage simulations, we use the $b-$values between $0$ and $3000$ $s/mm^2$.

\section{Results}
Fitting the above two models for the three directions diffusion weighting- $x,y,z$- we get the signal as a combination of the exponential decay curves for each of the directions as shown in the figures below. Also shown are the corresponding signal intensities $f_i$ corresponding to each of the ADCs. The three compartments - the Fiber compartment, the cell compartment, and the Free water compartment - in the voxel are first simulated separately, and the model is validated. Then the full structure voxel is simulated and validated. The results are shown in Figures \ref{fiberComp1}, \ref{cellComp1}, \ref{waterComp1} and \ref{fullStr1}.

\subsection{Key observations from $x,y,z$-directions diffusion weighted MC simulations}
Based on the basic MC simulations for the three directions diffusion weighting- $x,y,z$, we discover that the best fit radial diffusivity $\lambda_{\perp}$ comes out to be $0$. Observe that in Figure \ref{fiberComp1}, the signal decay in the $x$(and $y$)-direction is asymptotic to the signal intensity = 0.35, which is exactly the simulated fiber fraction. The corresponding ADC is made up primarily of $0$ (approx. $95 \%$, if we include nearby ADC of $0.2 \times 10^{-3} mm^2/s$); the remaining small fractions of higher ADCs can be explained by the model and simulation errors. As expected, for the cell compartment, the results shown in \ref{cellComp1} are almost identical, with the small variability attributable to the sampling variability of the random walks.The average ADC in all the three directions is approximately $0.2 \times 10^{-3} mm^2/s)$. For the water compartment, we get an average ADC of $2.33 \times 10^{-3} mm^2/s)$ in the $x$(and $y$)-directions; and it is almost $3\times 10^{-3} mm^2/s)$ in the $z$-direction. A different and higher ADC in free water in the $z$-direction is an important insight into the diffusion dynamics in the voxel. This clearly tells us that there is some anisotropy present in the EAEC compartment in addition to the anisotropy of the fiber compartment. This is also confirmed by the Signal decay curves and corresponding ADC spectrum in the full structure as shown in Figure \ref{fullStr1}.

Having validated the basic MC simulations and model, we simulate the axonal loss as follows. In the IA compartment, we simulate four cases where the fraction of axons with healthy axial diffusivity $\lambda_{\parallel}$, assumed to be $2 \times 10^{-3} mm^2/s$, is reduced progressively from $100 \%$ to $30 \%$. The unhealthy axons are assumed to have $\lambda_{\parallel}$ as $1 \times 10^{-3} mm^2/s$. The rest of the simulation parameters are the same as shown in table \ref{table:simParams}.

As detailed in the sections above, we first model the signal as in Eq. \ref{discrModel1} and solved using the $l_2$ minimization problem as in Eq. \ref{l2min}. This is done with different values of radial diffusivity $\lambda_{\perp}$ between $0$ and $0.4 \times 10^{-3} mm^2/s$, and finally the model with the least square residuals is selected. This first step in the DBSI-ADS method splits the signal into anisotropic and isotropic components. This model also gives us useful metrics - fiber fraction, cell fraction, best fit radial diffusivity $\lambda_{\perp}$. We run $10$ simulations with $1,000,000$ spins each. The results show a highly accurate recovery of the fiber and cell fractions. 

While the ground truth fiber fraction is $35 \%$, our model finds the fiber fraction very accurately. The One Sample t-test for fiber fraction gives the the fiber fraction as $34.83$, with 95 percent confidence interval: (34.10, 35.5) (P-value $<< 0.0001$).
Similarly for the extra-axonal cells, while the ground truth cell fraction is $35 \%$, our model finds the cell fraction very accurately too. The One Sample t-test for fiber fraction gives the the fiber fraction as $5.37$, with 95 percent confidence interval: (4.98, 5.75) (P-value $<< 0.0001$). The best fit radial diffusivity $\lambda_{\perp}$ comes out to be $0$, which also agrees with the results in the previous section. 

In the second step, the anisotropic component of the signal is isolated. From this anisotropic part of the signal we have to deduct a small percentage of signal at the higher end of the anisotropic ADC spectrum which is from the anisotropy from the EAEC compartment. After extensive simulations and validations, we arrive at $4\%$ as the fraction that we deduct. The anisotropic component recovers the fiber and cell fractions very accurately, but axial diffusivity spectrum has a percentage of signal around ADC of $3 \times 10^{-3} mm^2/s$, when the axial diffusivity simulated is $2 \times 10^{-3} mm^2/s$ or less. This percentage varies between $3\%$ and $6\%$ of the anisotropic fraction of $35\%$; so in the full signal this is between $1\%$ and $2\%$ of the complete signal. We also note here that, although the EAEC anisotropic fraction in our setup is minimal, this will increase as the fiber density increases and the inter-fiber distance shrinks in case of a pathology. We did simulate such cases and confirmed this increase in the EAEC anisotropic fraction. But for the setup that we simulate here, a minimally adjusted anisotropic component works well.

\subsection{Axonal Health}

\begin{figure}[!h]
\begin{center}
\includegraphics[width=2.2in]{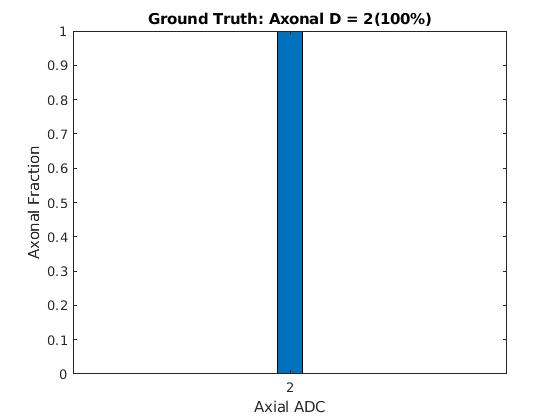}
\includegraphics[width=2.2in]{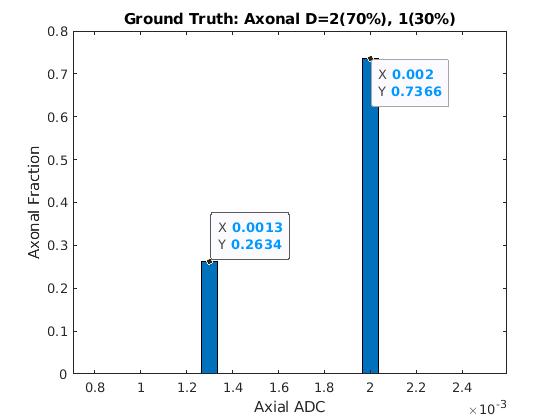}
\includegraphics[width=2.2in]{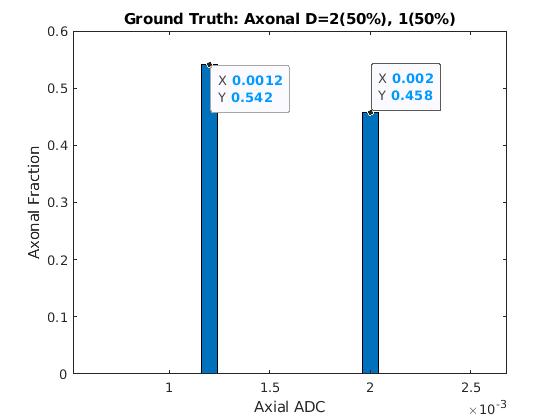}
\includegraphics[width=2.2in]{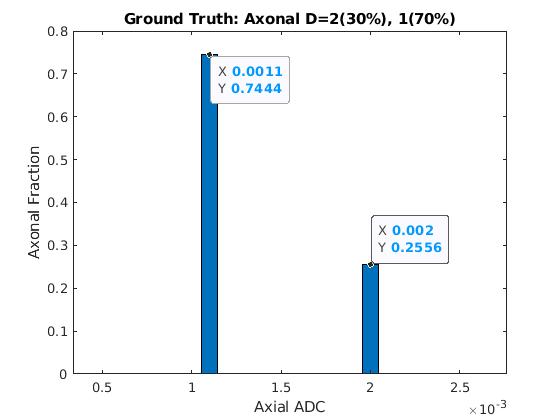}
\end{center}
\caption{Predicted fractions of healthy and unhealthy axons. Above each plot are the actual simulated fractions. The results are the average of $10$ sets for each simulation. The units for the Axial ADC are $mm^2/s$.}
\label{axialHealth_shaved_D2}
\end{figure} 

%
%
%
We model the net  anisotropic component of the signal as in Eq. \ref{dbsiADS_axonHealthEq} and solve as the minimization problem as in Eq. \ref{RASmin} for different values of unhealthy $\lambda_{\parallel}$, with $\lambda_{\parallel}=2 \times 10^{-3} mm^2/s$ for the healthy axons and $\lambda_{\perp}=0$. The model with the least BIC score is finally chosen. The results of the `Two component restricted anisotropic diffusion spectrum (RADS) model' for quantifying Axonal Health are shown in Figure \ref{axialHealth_shaved_D2}.

Our method produces highly accurate quantification of diseased and healthy axons with Pearson's correlation (predicted vs true proportion) $ r = 0.98 $ (p-value = 0.001); the one Sample t-test for proportion RMSE error gives the mean RMSE of 2\% (p-value = 0.034). Furthermore, the method finds the axial diffusivities of the diseased and healthy axons very accurately with mean error of 4\% (p-value = 0.001). For the Average vs predicted Average Axial ADC, we get the Pearson's correlation (predicted vs true proportion) $ r = 0.98 $ (p-value = 0.02). For the predicted Axial ADCs (predicted vs true ADCs), we get the Pearson's correlation (predicted vs true proportion) $ r = 0.99 $ (p-value $P < 0.0001$).

\section{Discussion}
In the present work, we develop an extension of DBSI (Diffusion Basis Spectrum Imaging) to quantify diseased axons in Multiple Sclerosis. The new method RADS-MRI models the anisotropic component of the diffusion- weighted signal as a linear combination of two parts - healthy and unhealthy axons based on the axial diffusivities. We run extensive MC simulations and validate our model. The first step of model recovers the fiber and cell fractions accurately with negligible variance, with Pearson’s correlation (predicted vs true proportion) of  $ r = 0.99 $ (p-value $< 0.001$).

Using Monte-Carlo (MC) simulations, we simulate different fractions of diseased and healthy axons. Our method produces highly accurate quantification of diseased and healthy axons with Pearson’s correlation (predicted vs true proportion) of  $ r = 0.99 $ (p-value = 0.001); the one Sample t-test for proportion error gives the mean error of 2\% (p-value = 0.034). Furthermore, the method finds the axial diffusivities of the diseased and healthy axons very accurately with mean error of 4\% (p-value = 0.001). The novelty of our method RADS comes from the fact that no other method uses axial diffusivity to distinguish healthy and unhealthy axons, and quantifies this reduction. Although various studies have shown axial diffusivity reduction to correlate with axonal damage (\cite{axial1}, \cite{SONG20031714}, \cite{KIM2006626}, \cite{SUN20061195}, \cite{SUN2}, \cite{KLAWITER20111454}, \cite{doi:10.1212/WNL.0b013e3181e042c4}, \cite{doi:10.1212/WNL.0b013e3181e0434d}), no method so far has been developed that uses this knowledge to quantify axonal damage.

There are some limitations of our research. The proposed new method has been validated on the synthetic simulated data only. Diffusion MRI has a low SNR, so the method needs to be further validated on real data. Although added Gaussian noise into our study, an in-vivo or ex-vivo examination is required to ascertain the applicability of this method to clinical setting. Furthermore, the ideal case analyzed here doesn't consider the case of crossing fibers. It must be noted that DBSI has been shown to accurately detect and quantify crossing fibers in animal models and human central nerve systems (\cite{Wang2011QuantificationOI},\cite{dbsiApp1}, \cite{dbsiApp2}). So, the crossing-fibers can be handled by our method by separately considering each fiber orientation and then quantifying fractions of diseased axons in each fiber orientation. So, although the theory is in place, the method still needs to be validated with in-vivo and ex-vivo cases of crossing fibers. 

Another assumption is that the healthy axonal anisotropic diffusion, taken to be $2 \times 10^{-3}mm^2/s$, is known. This also needs to be further validated with more experiments. This is critical especially because the model like ours has parameter identifiability problems, as these are what are called sloppy parameters (\cite{sloppyParams1}). In our model, if we don't constrain the healthy axial diffusivity as known, the model is generally unidentifiable, in the sense that many different combinations of healthy and unhealthy parts of the signal can produce the same combined signal. There is also a possibility that this healthy axial diffusion depends on other factors such as age, gender etc., which needs further research. We plan to address these limitations in our future work. 

In conclusion, RADS-MRI produces very accurate results to identify the proportion of healthy and diseased axons based on the axial diffusivities of axons on synthetic MC simulation data. Although this is encouraging, the method needs to be further validated with real data. If such examination succeeds, RADS-MRI derived reduction in axial diffusivity in axons has the potential to be used as a non-invasive biomarker for axonal damage in Multiple Sclerosis, and possibly other neuro-degenerative diseases.

\section*{Data Availability}
The code used will be shared upon reasonable request.

\bibliographystyle{unsrt} 
\bibliography{MC_axialFull_5.1.bib}
\end{document}